# Study on the Electronic Structure and Stability of Some OPE(oligo-phenylene-ethynylene derivative)-RE$_3$N@C$_{80}$ Dyads by PM7

Kye-Ryong Sin[1], Sun-Gyong Ko[1], Hong-Gol O[1], and Song-Jin Im[2]

[1]Department of Chemistry and [2]Department of Physics,
**Kim Il Sung** University, Daesong district,
Pyongyang, DPR Korea

E-mail: ryongnam9@yahoo.com

**Abstract**: In this paper, we investigated the electronic structure and stability of some mesomorphic OPE-RE$_3$N@C$_{80}$ dyads from the oligo-phenylene-ethynylene derivatives (OPE) and the trimetallic nitride template endohedral fullerenes (TNT-EMFs) - RE$_3$N@C$_{80}$ (RE=Sc,Y,La) by using PM7, the updated version of the semi-empirical Hartree-Fock method. In OPE-RE$_3$N@C$_{80}$, the fullerene cages were modified to have the opened cage (fulleroid) structure by addition of OPE on the [6,6] position of the fullerene cages. There was no considerable charge transfer between OPE and fullerene cage, but the fullerene cages had the remarkable minus charges mainly due to the electron transfer from RE$_3$N to the cage. The calculated electronic spectra showed that light absorption bands of OPE-C$_{80}$ were more red-shifted than that of OPE-RE$_3$N@C$_{80}$ and all of OPE-RE$_3$N@C$_{80}$ seem to have a couple of Vis-NIR absorption peaks.

## 1. Introduction

Now it is well-known that fullerenes are not chemically inert and undergo various chemical reactions such as nucleophilic addition, Diel's-Alder reaction, 1,3-dipolar cycloaddition, radical addition, oxidation, reduction etc.[1,2,3]

In these decades many researches in the fullerenes chemistry have been focused on the synthesis of the endohedral metallofullerenes (EMFs), the so-called "cluster-fullerene", containing metal atoms or clusters therein and on their application in manufacture of the novel nano-materials such as molecular devices, sensors and medical tools[4]. Especially, a variety of the trimetallic nitride template endohedral fullerenes (TNT-EMFs) such as RE$_3$N@C$_n$ ( RE = Sc, Y, La; n=78, 80, 82, ⋯ ) have been synthesized and modified for utilizing their functionalities in molecular electronics and bio-technology.[5,6] One of the most prosperous applications of $RE_3N@C_n$ can be found in organic photovoltaics due to their excellent electron acceptor abilities. A recent research was carried out for synthesis of some π-conjugated system – fullerene dyads for photovoltaic applications, where the donor units were either oligo-phenylene-ethynylene (OPE) or oligo-phenylene-vinylene (OPV) derivatives and for the acceptor, $C_{60}$ or $Y_3N@C_{80}$ were used.[7] The liquid crystalline (LC) behavior, shown by the synthesized dyads was expected to improve the photovoltaic efficiency of the BHJ (block hetero-junction) organic solar cells by ambipolar charge transfer.

In this paper, PM7 in MOPAC2012[8], one of the updated semi-empirical Hartree-Fock methods, was applied in the theoretical study on the electronic structure and stability of OPE-$C_{80}$ and OPE-$RE_3N@C_{80}$ dyads (RE = Sc, Y, La). There have been some reports on DFT (Density Functional Theory) study on $RE_3N@C_n$ and their derivatives[4,9,10,11], but still no research has been done on theoretical calculation of the electronic structures of the OPE-$RE_3N@C_{80}$ dyads.

## 2. Computational Models and Method

Here the quantum chemical study has been done on four OPE-FD dyads (OPE-$C_{80}$ and OPE-$RE_3N@C_{80}$), where FD means $C_{80}$ and three kinds of $RE_3N@C_{80}$ (RE=Sc,Y,La).

For all the OPE-FDs, the geometry of $C_{80}$-$I_h$, one of the geometric isomers of $C_{80}$, was chosen as the fullerene cage, where $I_h$ shows the geometric symmetry of the fullerene cage. (Figure 1)

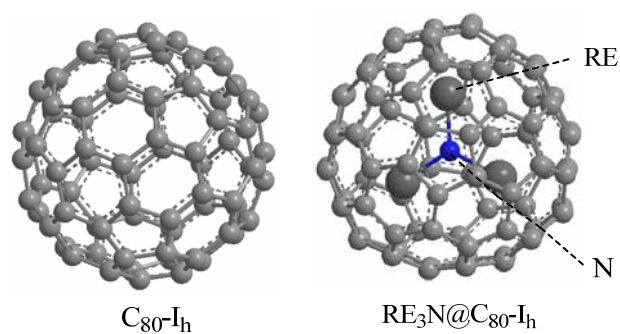

$C_{80}$-$I_h$      $RE_3N@C_{80}$-$I_h$

Figure 1. Models for FDs (RE=Sc,Y,La)

Figure 2 shows the chemical structures of the OPE-$Y_3N@C_{80}$ dyad, synthesized in the previous research.[7] According to that experimental research, the models for OPE-$C_{80}$ and OPE-$RE_3N@C_{80}$ dyad (RE=Sc, Y, La) were chosen as [6,6] adducts, where OPE was covalently linked to $C_{80}$ or $RE_3N@C_{80}$ just on the [6,6] addition site of the fullerene cage (the nearest site to RE atom in case of $RE_3N@C_{80}$). To simplify the task and avoid the overload in computation, the long alkyl chain R (-$C_{12}H_{25}$) in the OPE was shortened as -$CH_3$ in all the models for OPE-FDs.

The OPE-FDs can be separated as two individual subunits, OPE and FD, for comparing their electron donor – acceptor interaction. Here $OPE_1$ symbolizes a half part of OPE (Figure 3).

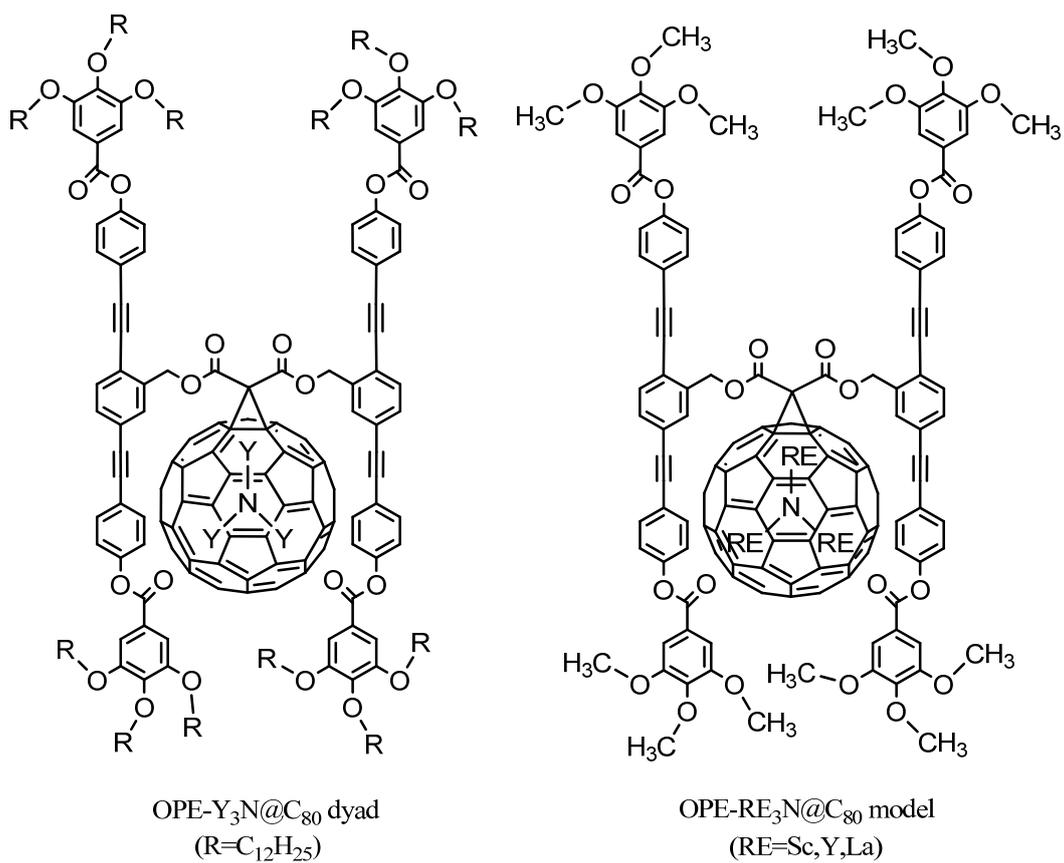

OPE-Y₃N@C₈₀ dyad
(R=C₁₂H₂₅)

OPE-RE₃N@C₈₀ model
(RE=Sc,Y,La)

Figure 2. Models for OPE-FDs

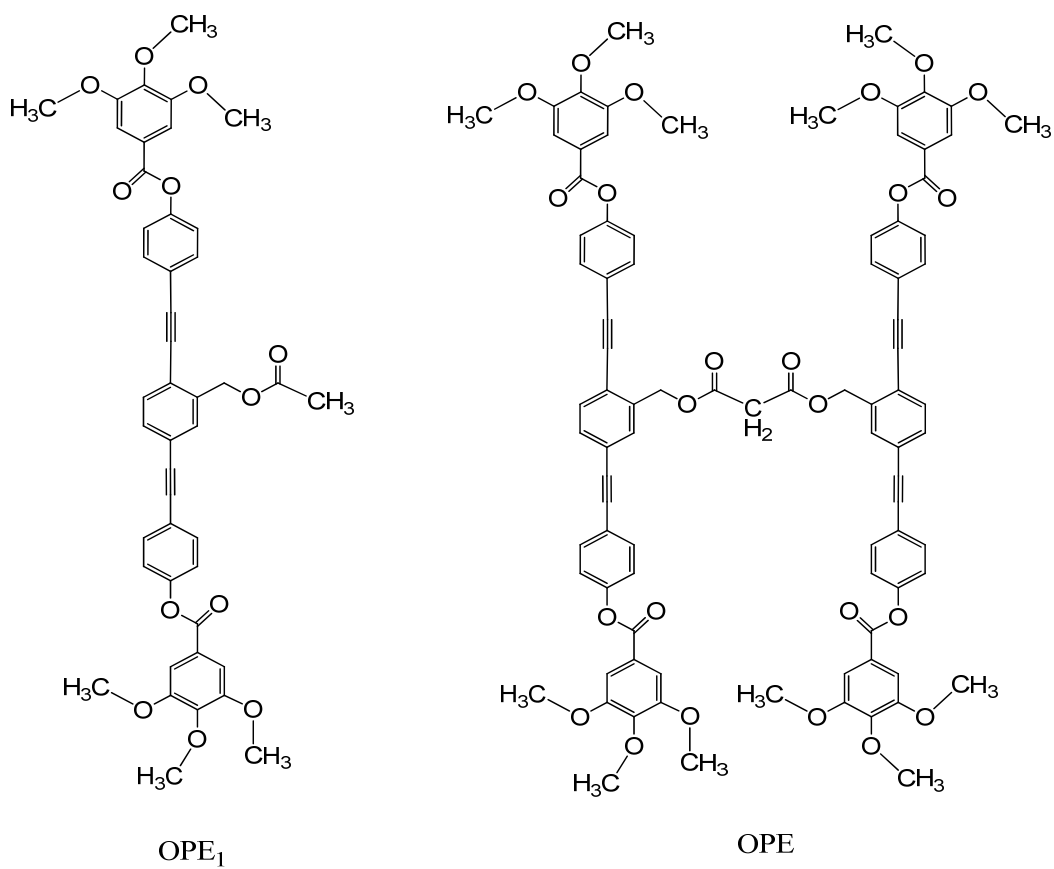

OPE₁

OPE

Figure 3. Models for OPE and OPE₁

To consider the effect of $RE_3N$ on the electronic structure of the OPE-$RE_3N$@$C_{80}$, the empty fullerene cage model without $RE_3N$ was also calculated as OPE-$C_{80}$.

The geometric and electronic structures of the models were calculated by PM7 from **MOPAC2012**, the latest version of the semi-empirical MO software package, that has been well-known as one of the most efficient quantum chemical tools with the enhanced accuracy for a wide range of molecules, complexes, polymers, crystals, and TNT-EMFs.[8,12] Especially, it offers good parameter set for the calculation of most of the elements on the periodic table including rare earth elements.

The first step of calculation was the geometry optimization by EF (Eigenvector-Following) routine, which was followed by the configuration interaction (CI) calculation based on the single-point MO results. The configurations for the singlet electronic transitions were composed of 20 MOs near HOMO and LUMO (10 occupied MOs and 10 unoccupied MOs). The electronic spectra were drawn by using the Gaussian smoothing function based on the transition energies (mode positions) and the oscillator strengths (mode intensities).

Relative Stability of the OPE-FDs was evaluated by $\Delta E_t$, the difference of total energies ($E_t$) between the resulting model (OPE-FD) and the separated subunits (OPE and FD).

### 3. Results and Discussion

**1) The geometric and electronic structures of the separated subunits (OPE and FDs)**

Figure 4 shows the optimized geometric structures of OPE and its one branch ($OPE_1$), where the phenylethynyl unit (-$C_6H_4$-C≡C-$C_6H_4$-C≡C-$C_6H_4$-) is arranged to forms a straight line, but its three phenyl rings are not on the same plane, which prevents to form larger π-conjugation plane in OPE.

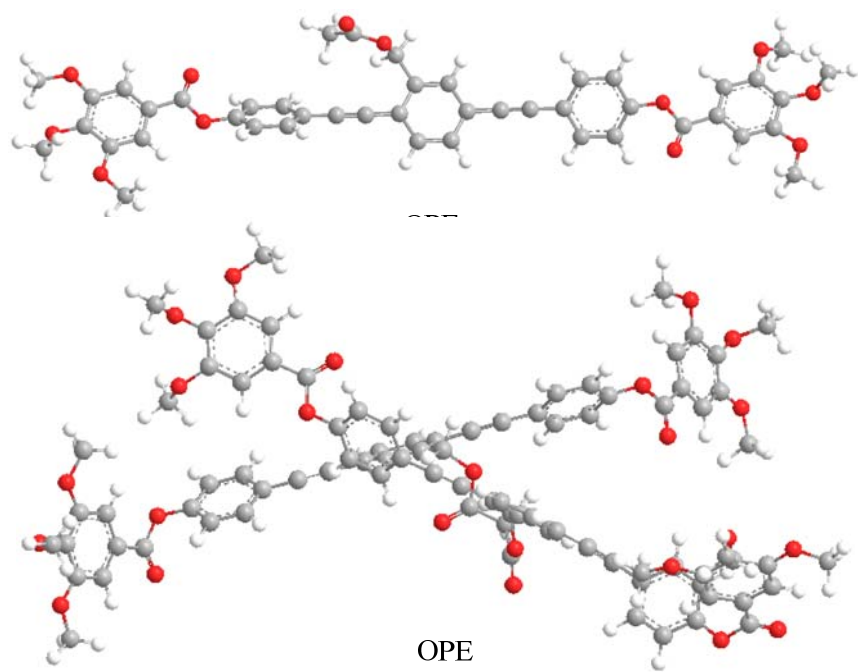

Figure 4. The optimized geometric structures of OPE and its one branch ($OPE_1$)

It can be seen from the optimized geometric structures of RE$_3$N@C$_{80}$ in Figure 5 that RE$_3$N (RE = Sc, La) has the planar form, but Y$_3$N has the pyramidal form, which resembles the previous XRD measurement of Gd$_3$N@C$_{80}$-I$_h$ and DFT calculation of Y$_3$N@ C$_{78}$- D$_{3h}$. [13,14]

From the electronic structures of RE$_3$N@C$_{80}$ calculated from their optimized geometry, it was found out that the positive charge of RE atoms was increased and the negative charge of N atom decreased in the cluster fullerene compared with those in free RE$_3$N, which shows that in RE$_3$N@C$_{80}$ more portion of electrons of RE atoms was transferred to the fullerene cage, not to N atom.[12]

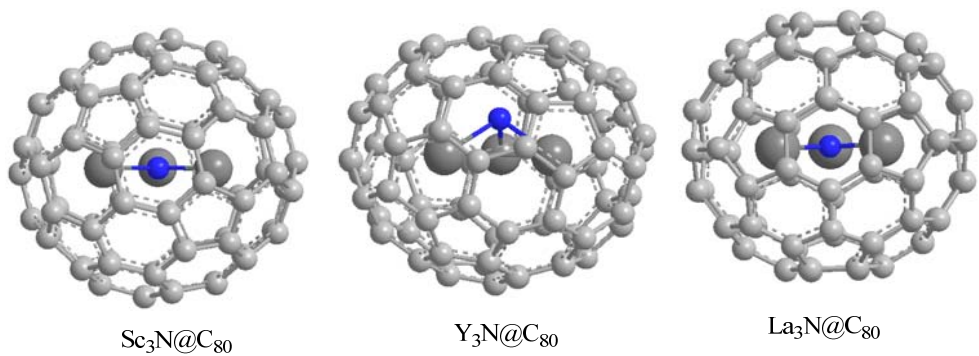

Sc$_3$N@C$_{80}$    Y$_3$N@C$_{80}$    La$_3$N@C$_{80}$

Figure 5. The optimized geometric structures of RE$_3$N@C$_{80}$

Figure 6 shows that OPE$_1$ and OPE have the similar HOMO-LUMO levels, which means there can not be apparent π-conjugation between two phenylethynyl branches in OPE. All FDs (C$_{80}$ and RE$_3$N@C$_{80}$) have lower LUMO levels than OPE, therefore they can accept electron from OPE. HOMO levels of RE$_3$N@C$_{80}$ are lower than that of the empty C$_{80}$. It can be explained as the result of stabilization of the C$_{80}$ cage by RE$_3$N incorporation.[15]

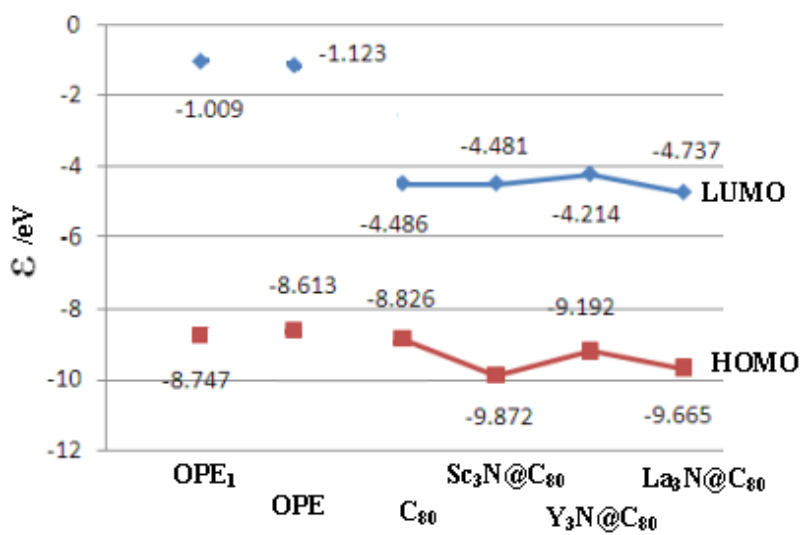

Figure 6. HOMO – LUMO energy levels of subunits of OPE-FDs

From Figure 7, it can be seen that OPE has UV absorption, but FDs can interact with visible light or even with NIR.

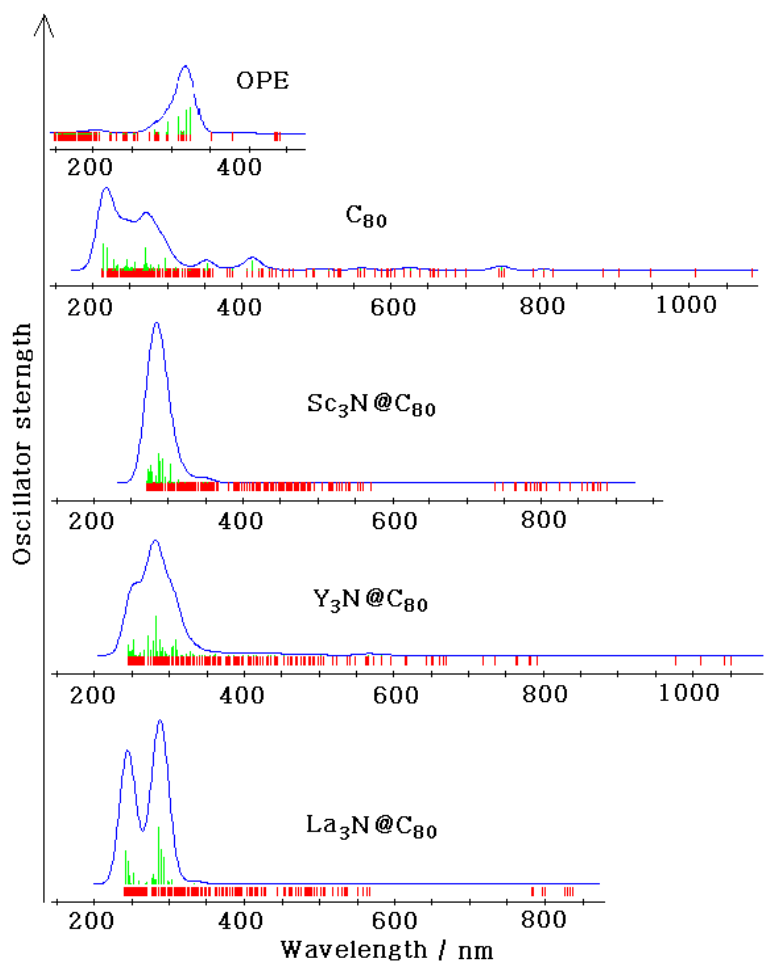

Figure 7.　The electronic spectra of OPE and FDs

( red: mode positions, green: mode intensities )

**2) The geometric and electronic structures of OPE-FDs**

Four models of the OPE-FDs discussed in this paper had the similar configurations after geometric optimization (Figure 8).

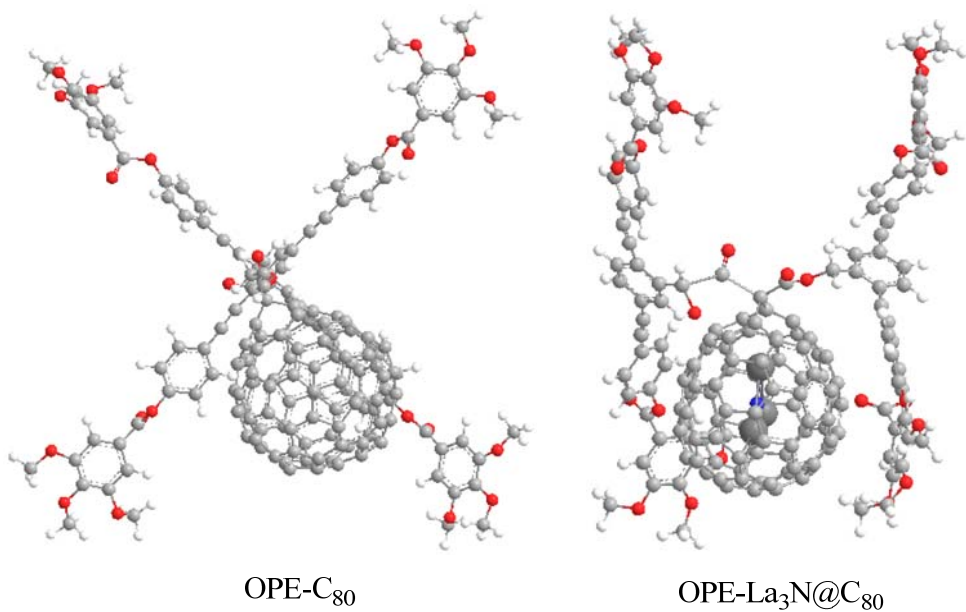

OPE-$C_{80}$　　　　　　　　　　OPE-$La_3N@C_{80}$

Figure 8. The different views of the optimized structures of OPE-FDs

These configurations may be different from those of the real OPE-FD dyads[7] because these models have the shorten alkyl group (-CH$_3$) instead of the long chain (–C$_{12}$H$_{25}$) and can not express their well-assembled frameworks in the liquid crystalline phase.

Figure 9 shows $\Delta E_t$ of the OPE-fullerenes calculated as the total energy difference of OPE-FD from its separated subunits (OPE and FD). OPE-C$_{80}$ became to be unstable after the formation of the dyad and it can be considered as the result of structural deformation of the subunits, especially C$_{80}$ due to the formation of the dyad. The most stable one was OPE-La$_3$N@C$_{80}$ and other OPE-RE$_3$N@C$_{80}$ were also more stable than OPE-C$_{80}$ because all of RE$_3$N@C$_{80}$ had been stabilized by electron transfer from RE$_3$N to C$_{80}$.

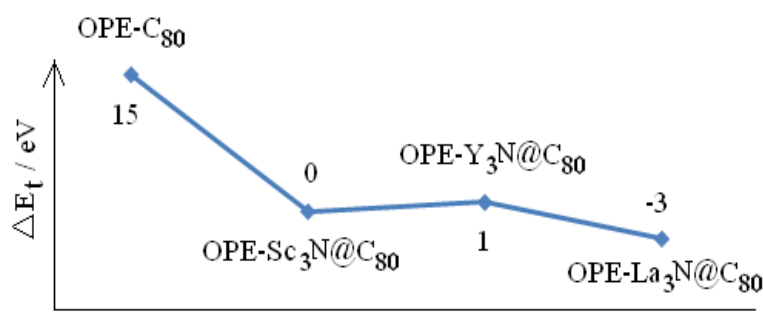

Figure 9. Relative stability ($\Delta E_t$) of OPE-FDs

In all of OPE-FDs, the fullerene cages were modified to have the open-up cage (fulleroid) structures. Table 1 shows the C-C distances at the [6,6] OPE-addition sites of FDs and their increases ($\triangle R_{C-C}$) after OPE-addition, where more stable OPE-La$_3$N@C$_{80}$ had the less $\triangle R_{C-C}$ and less stable OPE-C$_{80}$ and OPE-Y$_3$N@C$_{80}$ had the larger $\triangle R_{C-C}$.

Table 1.    C-C distance at the [6,6] addition site of FDs (nm)

| model | OPE-C$_{80}$ | OPE-Sc$_3$N@C$_{80}$ | OPE-Y$_3$N@C$_{80}$ | OPE-La$_3$N@C$_{80}$ |
|---|---|---|---|---|
| free FD | 0.142 | 0.148 | 0.153 | 0.149 |
| OPE-FD | 0.235 | 0.239 | 0.247 | 0.234 |
| $\triangle R_{C-C}$ | 0.093 | 0.091 | 0.094 | 0.085 |

Table 2 shows the local charges of the subunits (OPE, C$_{80}$ cage, RE$_3$N) in OPE-FDs, where there was no considerable charge transfer between OPE and FDs, but in OPE-RE$_3$N@C$_{80}$ the fullerene cages had the remarkable minus charges mainly due to the electron transfer from RE$_3$N to the cage. It seems that the more electrons were transferred from RE$_3$N to C$_{80}$, the more stable OPE-RE$_3$N@C$_{80}$ was formed.

Table 2.    Local charges of the subunits in OPE-FDs (e)

| model | OPE-C$_{80}$ | OPE-Sc$_3$N@C$_{80}$ | OPE-Y$_3$N@C$_{80}$ | OPE-La$_3$N@C$_{80}$ |
|---|---|---|---|---|
| OPE | 0.012 | 0.071 | 0.037 | 0.069 |
| C$_{80}$ cage | -0.012 | -4.454 | -3.976 | -4.471 |
| RE$_3$N | - | 4.383 | 3.939 | 4.402 |

From the electronic spectra of OPE-FDs (Figure 10), it can be found out that light absorption band of OPE-$C_{80}$ was more red-shifted than that of OPE-$RE_3N@C_{80}$, but its maximum absorption intensity was far less than OPE-$RE_3N@C_{80}$, and all of OPE-FDs seem to have a couple of Vis-NIR absorption peaks.

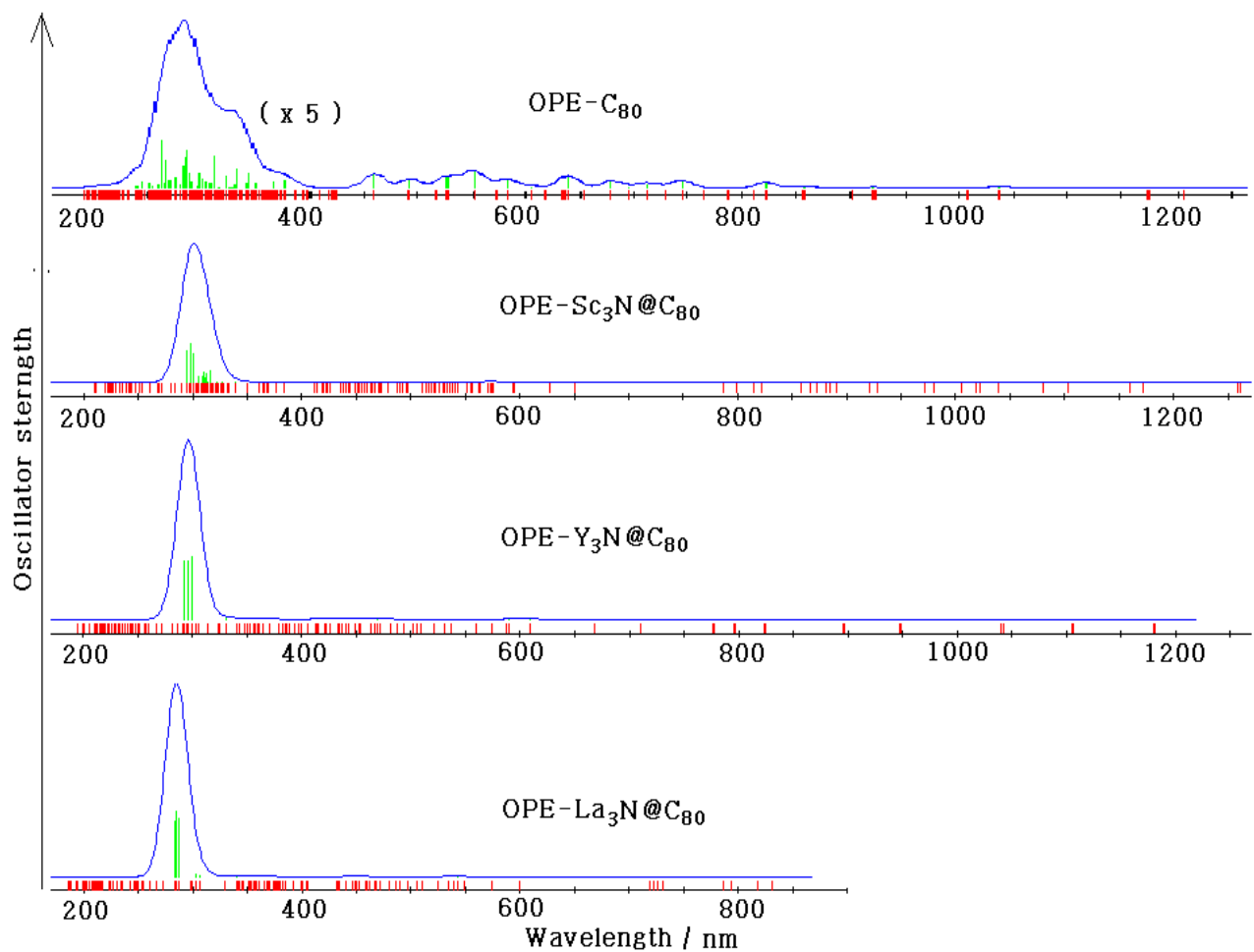

Figure 10.　The electronic spectra of OPE-FDs
( red: mode positions, green: mode intensities )

## 4. Conclusion

PM7 calculations were carried out on four OPE-fullerene dyads ( OPE-FDs ) such as OPE-$C_{80}$ and OPE-$RE_3N@C_{80}$ ( RE = Sc, Y, La ).

In all of OPE-FDs, the fullerene cages were modified to have the open-up cage (fulleroid) structure by addition of OPE on the [6,6] position of the fullerene cages. The C-C distance at the [6,6] addition site of the cages was less increased in the more stable OPE-FDs.

There was no considerable charge transfer between OPE and FDs, but in OPE-$RE_3N@C_{80}$ the fullerene cages had the remarkable minus charges mainly due to the electron transfer from $RE_3N$ to the cage.

Light absorption bands of OPE-$C_{80}$ were more red-shifted than that of OPE-$RE_3N@C_{80}$ and all of OPE-FDs seem to have a couple of Vis-NIR absorption peaks.

**Acknowledgement**

The authors thank Stewart Computational Chemistry for its efficient **MOPAC2012**.